\documentclass{article}  
\usepackage{spadre2008}
\usepackage{graphicx}
\frompage{000} \topage{000}                                              
\def\rightmark{Azimuthal Correlations with High-$p_{T}$ Multi-hadron 
Cluster Triggers}
\def\leftmark{B. Haag}
 
\title{Azimuthal Correlations with High-$p_{T}$ Multi-hadron 
Cluster Triggers in Au+Au Collisions at $\sqrt{s_{NN}} = 200$ GeV from STAR} 
\authors{
{B. Haag$^1$ for the STAR Collaboration %
}\\[2.812mm]
{\normalsize
\hspace*{-8pt}$^1$ University of California, Davis, \\ 
95616 Davis, USA\\[0.2ex] 
}}
 
\abstract{Di-hadron correlation measurements have been used to probe di-jet
production in collisions at RHIC. A strong suppression of the
away-side high-$p_{T}$ yield in these measurements is direct evidence that
high-$p_{T}$ partons lose energy as they traverse the strongly interacting
medium.  However,
since the momentum of the trigger particle is not a good measure of the jet
energy, azimuthal di-hadron correlations have limited sensitivity to the
shape of the fragmentation function.  We explore the possibility to
better constrain the initial parton energy by using clusters of multiple
high-$p_{T}$ hadrons in a narrow cone as the `trigger particle' in the azimuthal
correlation analysis.  We present first results from this analysis of
multi-hadron triggered correlated yields in Au+Au collisions at
$\sqrt{s_{NN}}$ = 200 GeV from STAR. The results are compared to
Pythia calculations, and the
implications for energy loss and jet fragmentation are discussed.}

 
\begin{document}
 
\maketitle
\setcounter{page}{1}

\section{Introduction}\label{intro}

Recent experiments at RHIC have shown that
in high-energy heavy-ion collisions, a strongly coupled
medium consisting of deconfined quarks and gluons has been
produced \cite{Adams:2005dq}.
This medium is opaque to hard scattered partons, which lose energy
as they traverse the medium and subsequently
their fragmentation is modified \cite{Adler:2002xw}. 
This fragmentation has been studied using azimuthal
correlations of hadrons with large transverse momentum ($p_{T}$).

Due to the large particle multiplicities observed in heavy-ion collisions, our
current method of measuring jet-like correlations is via di-hadron correlations
\cite{Adams:2006yt}.
To characterise parton energy loss, we would like to measure the 
fragment distribution of hadrons in jets. So far, di-hadron correlations 
have been used for this purpose, since the large background of soft 
particles produced in heavy ion collisions makes it difficult to 
directly reconstruct jets. In these measurements, the transverse 
momentum of a trigger particle, $p_{T}^{trig}$, is used as a proxy for the jet 
energy, $E_{T}^{jet}$. In this paper, we present a new method, using a cluster of 
multiple high-$p_{T}$ hadrons as a trigger. Multi-hadron clusters may provide 
a better measure of the jet energy than than the leading particle $p_{T}$.

\section{Experimental Setup}\label{techno}  

 \begin{figure}[!b]
\centering
\includegraphics[width=1.0\textwidth]{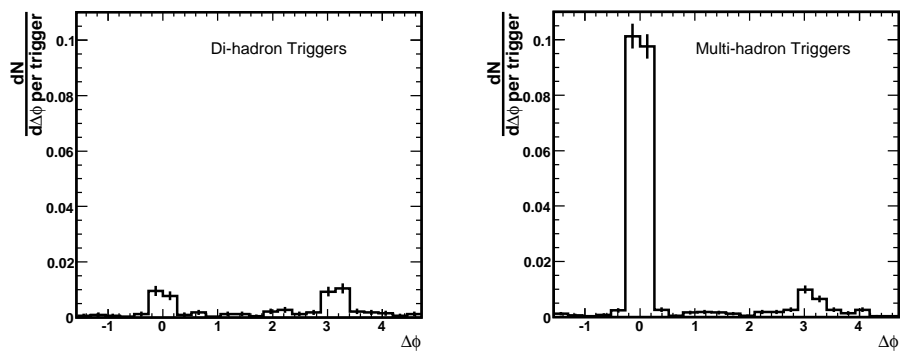}
\caption{Background subtracted azimuthal distributions 
for di-hadron triggers (left) and multi-hadron triggers (right) for 
$12 < p_{T}^{trig} < 15$ GeV/$c$ and 
4.0 GeV/$c$ $<$ $p_{T}^{assoc}$ $<$ 5.0 GeV/$c$.  A minimum
secondary seed of 3.0 GeV/$c$ is used.}
\label{dNdDphi}
\end{figure}

There are approximately 24M Au+Au events at $\sqrt{s_{NN}} = 200$ GeV
used in this study.  They are taken from the data collected
during the year 4 run at RHIC, from the 0-12\% most central 
events, selected via STAR's Zero Degree Calorimeters. 
Details on the triggering and particle reconstruction are
discussed elsewhere \cite{Ackermann:2002ad}.

\section{Analysis and discussion}

Charged tracks from primary vertices are used to construct 
multi-hadron and di-hadron azimuthal distributions.  
The tracks are selected within the pseudo-rapidity range
of  $|\eta| < 1$.  The uncorrelated background is removed
using the zero yield at minimum (ZYAM) \cite{Ajitanand:2005jj} 
method.  As elliptic flow ($v_{2}$) is less than a 1\% modulation of the background
in the ranges selected for $p_{T}^{trig}$ and $p_{T}^{assoc}$
and the signal to background is much larger than 1\%, 
the elliptic flow modulation is considered negligible in this analysis.
 
 \begin{figure}[!b]
\centering
\includegraphics[width=1.0\textwidth]{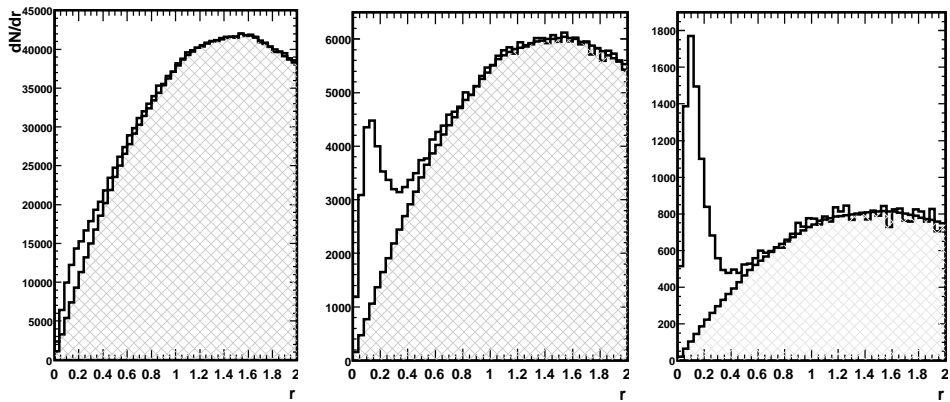}
\caption{Radial distributions of triggers with associated tracks
from the same event (white histograms) 
and from different events (hatched histograms).
Panels from left to right show minimum secondary seed
cuts of 2.0, 3.0, and 4.0 GeV/$c$ respectively.}
\label{randomClusters}
\end{figure}
 
When forming multi-hadron triggers, all 
tracks which pass the track quality cuts with $p_{T} > 5.0$ GeV/$c$
are collected as ``primary seeds''.  Then within a cone radius
($r = \sqrt{\Delta\phi^{2}+\Delta\eta^{2}}$) of 0.3,
all ``secondary seeds''
which fall above a minimum $p_{T}$ cut are collected.  Minimum 
secondary seed cuts of 2,3, and 4
GeV/$c$ have been used for a systematic study.  
Next, the sum of the primary and secondary seeds is taken to 
be the trigger $p_{T}$.  To illustrate, a multi-hadron trigger of 12 GeV/$c$ might be a combination
of a 6 GeV/$c$ primary seed and two secondary seeds of 3 and 3 GeV/$c$ each, while in the
standard di-hadron analysis \cite{Adler:2002tq}, the trigger would be a single hadron 
with $p_{T}$ = 12 GeV/$c$.
With the multi-hadron triggers defined, azimuthal difference distributions are 
calculated between the primary seed in the cone and associated tracks with $p_{T}$ greater than
the minimum secondary seed $p_{T}$ cut.  Representative distributions are shown in Figure \ref{dNdDphi}.
For the multi-hadron triggers there is a bias on the near-side
due to the algorithm which artifically enhances the yield.
With these distributions, recoil (away-side) yields are extracted and studied for various $p_{T}^{trig}$ bins.  
 
Random combinations occur in the multi-hadron cluster algorithm.
The multi-hadron clusters contain a combinatorial background in which a seed particle 
from a jet is combined with one or more secondary seeds from the 
underlying soft event.  To study this background,
the radial distributions of primary seeds for two different cases are constructed:
with associated tracks in the same event and with associated tracks in 
different events.  These distributions are shown in Figure \ref{randomClusters}
with the open histograms showing same event correlations and 
the grey filled histograms showing correlations from mixed events, taking the seed track and the 
secondary seeds from different events.  The background histograms have been scaled to the signal
histograms.  The secondary seed $p_{T}$ increases from left to right
with $p_{T} >$ 2.0, 3.0, and 4.0 GeV/$c$ and the signal-to-background
increases from 0.2 to 0.7 to 2.0 respectively.
A radius of 0.3 along with a minimum secondary seed $p_{T}$ cut
greater than 3.0 GeV/$c$ leads to a reasonable signal to background for this study.
Future plans include background subtracted yields calculated with an 
estimate of background trigger yields.

\begin{figure}[!hb]
\centering
\includegraphics[scale=1.0]{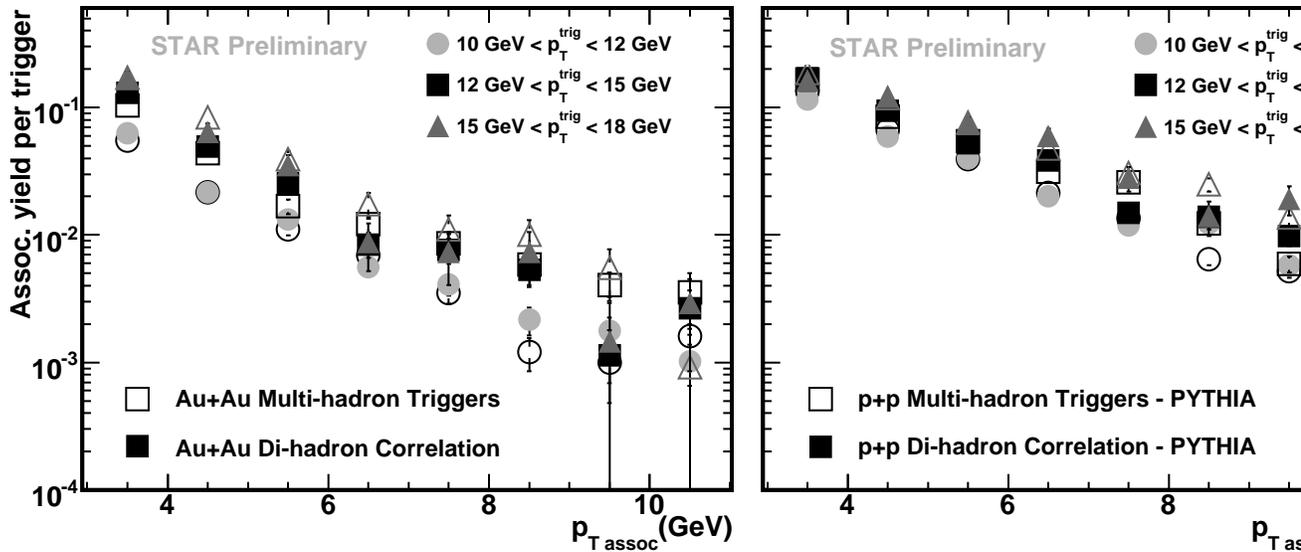}
\caption{Recoil yield per trigger for three $p_{T}$ bins:
$10 < p_{T}^{trig} < 12$ GeV/$c$ (circles),
$12 < p_{T}^{trig} < 15$ GeV/$c$ (squares),
and $15 < p_{T}^{trig} < 18$ GeV/$c$ (triangles).
Data is presented on the left (Au+Au), Pythia predictions
are presented on the right (p+p).
A minimum secondary seed cut of $p_{T} >$ 3.0 GeV/$c$
is applied.}
\label{away_yields_30}
\end{figure}

\begin{figure}[!t]
\centering
\includegraphics[scale=1.0]{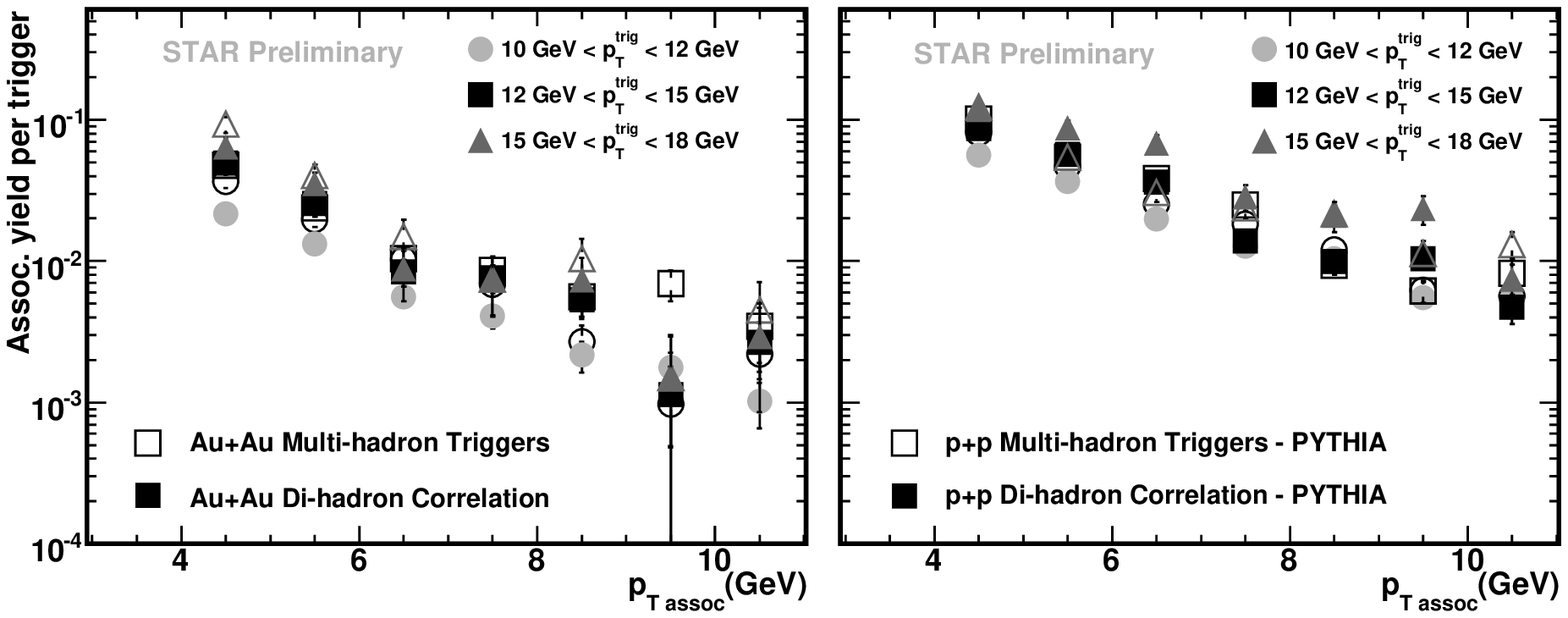}
\caption{Recoil yield per trigger for three $p_{T}$ bins:
$10 < p_{T}^{trig} < 12$ GeV/$c$ (circles),
$12 < p_{T}^{trig} < 15$ GeV/$c$ (squares),
and $15 < p_{T}^{trig} < 18$ GeV/$c$ (triangles).
Data is presented on the left (Au+Au), Pythia predictions
are presented on the right (p+p).
A minimum secondary seed cut of $p_{T} >$ 4.0 GeV/$c$
is applied.}
\label{away_yields_40}
\end{figure}


Figures \ref{away_yields_30} and \ref{away_yields_40}
show recoil (away-side) yields for three $p_{T}$ bins:  $10 < p_{T}^{trig} < 12$ GeV/$c$, 
$12 < p_{T}^{trig} < 15$ GeV/$c$, and $15 < p_{T}^{trig} < 18$ GeV/$c$ respectively.
Figure \ref{away_yields_30} shows a comparison between
multi-hadron (open symbols) and di-hadron (solid symbols) 
triggers with a minimum secondary
seed cut of $3.0$ GeV/$c$ for the data (left panels)
and Pythia (right panels).
The same comparisons are shown in Figure \ref{away_yields_40} but for a higher
minimum secondary seed cut of $4.0$ GeV/$c$.


The associated per-trigger yields for both single-hadron and multi-hadron triggers
in Figures \ref{away_yields_30} and \ref{away_yields_40} 
are similar, suggesting the selection of a similar underlying jet-energy distribution by 
both methods.  The same analysis performed on Pythia events is shown in the right-hand
panels of Figures \ref{away_yields_30} and \ref{away_yields_40}.
In the Pythia events, the di-hadron analysis and multi-hadron triggered analysis also 
give similar results, although the per-trigger yields
are generally higher than measured in the data.    

\section{Conclusions}\label{concl}
This paper has presented first results on the use of multi-hadron triggers 
investigated as the next step toward full jet reconstruction in heavy-ion collisions.  
A cone radius of 0.3 coupled with a minimum secondary seed cut greater than 3.0 GeV/$c$ leads 
to a reasonable signal to background ratio of 0.7.  
Moreover, the away-side yields for multi-hadron correlations 
and from di-hadron measurements are consistent.
This effect is also reproduced in Pythia simulations.
Further analysis of the Pythia events
to compare the underlying jet energy selections for di-hadron analysis
and multi-hadron triggered analysis is ongoing.

\vfill\eject
\end{document}